\documentclass[aps,prl,twocolumn,preprintnumbers,superscriptaddress
]{revtex4-2}
\usepackage{graphicx}
\usepackage[utf8]{inputenc}
\usepackage[T1]{fontenc}
\usepackage{xcolor}
\usepackage{textcomp}
\usepackage{bm}
\usepackage{siunitx}
\usepackage{physics}
\usepackage{amsmath}
\usepackage{tikz}
\usepackage{mathdots}
\usepackage{yhmath}
\usepackage{cancel}
\usepackage{color}
\usepackage{array}
\usepackage{multirow}
\usepackage{amssymb}
\usepackage{gensymb}
\usepackage{tabularx}
\usepackage{extarrows}
\usepackage{booktabs}
\usetikzlibrary{fadings}
\usetikzlibrary{patterns}
\usetikzlibrary{shadows.blur}
\usetikzlibrary{shapes}
\usepackage{makecell}
\usepackage{float}

\usepackage{color}

\definecolor{LinkColor}{rgb}{0.0,0.0,1.0}

\usepackage{hyperref}
\hypersetup{
	colorlinks=true,
	citecolor=LinkColor,
	linkcolor=LinkColor,
	urlcolor=LinkColor,
}

\usepackage{listings}
\definecolor{lightgray}{gray}{1}

\begin{document}

\title{Fractional quantum anomalous Hall and anyon density-wave halo \\in a minimal interacting lattice model of twisted bilayer MoTe$_2$}

\author{Chuyi Tuo}
\affiliation{Institute for Advanced Study, Tsinghua University, Beijing 100084, China}
\author{Ming-Rui Li}
\affiliation{Institute for Advanced Study, Tsinghua University, Beijing 100084, China}
\author{Hong Yao}
\email{yaohong@tsinghua.edu.cn}
\affiliation{Institute for Advanced Study, Tsinghua University, Beijing 100084, China}

\date{\today}

\begin{abstract}
The experimental discovery of fractional quantum anomalous Hall (FQAH) states in tunable moiré superlattices has sparked intense interest in exploring the interplay between topological order and symmetry breaking phases. 
In this paper, we present a comprehensive numerical study of this interplay through large-scale density matrix renormalization group (DMRG) simulations on a minimal two-band lattice model of twisted bilayer MoTe$_2$ at filling $\nu=-2/3$. 
We find robust FQAH ground states and provide clear numerical evidences for anyon excitations with fractional charge and pronounced real-space density modulations, directly supporting the recently proposed anyon density-wave halo picture.
We also map out the displacement field dependent phase diagram, uncovering a rich landscape of charge ordered states emerging from the FQAH, including a quantum anomalous Hall crystal (QAHC) with an integer quantized Hall conductance.
We expect our work to inspire further research interest of intertwined correlated topological phases in moiré systems.
\end{abstract}

\maketitle

\textit{Introduction.}---The fractional quantum Hall effect (FQHE)~\cite{tsui1982two,laughlin1983anomalous,PhysRevLett.51.605,PhysRevLett.63.199,PhysRevB.46.2223} represents one of the most profound phenomena in modern condensed matter physics, offering a canonical example of intrinsic topological order~\cite{wen1990topological} characterized by fractionalized anyon excitations~\cite{laughlin1983anomalous,arovas1984fractional}, topological ground state degeneracy~\cite{wen1990ground}, emergent gauge field~\cite{zhang1989effective}, and long-range quantum entanglement~\cite{kitaev2006topological,levin2006detecting}. While traditional studies of the FQHE have largely focused on two-dimensional electron gases under strong magnetic fields, its zero field analog on lattice systems, the FQAH state, has long been theoretically anticipated to emerge in bands with nontrivial Chern numbers~\cite{tang2011high,sun2011nearly,neupert2011fci,sheng2011fractional,regnault2011fci,bergholtz2013topological,PhysRevLett.108.126805,PhysRevB.86.201101} and favorable quantum geometry~\cite{PhysRevB.90.165139,PhysRevResearch.2.023237,wang2021exact,ledwith2023vortexability,estienne2023ideal,liu2025theory}. Very recently, breakthrough in moiré van der Waals heterostructures~\cite{bistritzer2011moire,wu2018hubbard,PhysRevLett.122.086402,andrei2020graphene,andrei2021marvels,kennes2021moire,castellanos2022van,mak2022semiconductor,nuckolls2024microscopic} have enabled the experimental realization of FQAH in various experimental systems. Following initial signatures in twisted bilayer graphene under a small magnetic field~\cite{xie2021fractional}, robust zero-field FQAH have been decisively observed in twisted bilayer MoTe$_2$ ($t\text{MoTe}_2$)~\cite{cai2023signatures,zeng2023thermodynamic,park2023observation,xu2023observation,Redekop2024,ji2024local,xu2025signatures} and in aligned multilayer rhombohedral graphene/hBN~\cite{Lu2024fractional,xie2024even,choi2024electricfieldcontrolsuperconductivity,WatersChern2025,75gl-jzl6,lu2025extended,zhou2024layer,han2024engineering,ding2024electricalswitchingchiralityrhombohedral,zheng2024switchablecherninsulatorisospin,xiang2025continuouslytunableanomaloushall,wang2025electricalswitchingcherninsulators,li2025tunablechern}, sparking a surge of interest in the microscopic mechanisms governing these correlated topological phases~\cite{li2021spontaneous,PhysRevB.111.125122,PhysRevB.108.085117,PhysRevLett.132.036501,xu2024maximally,PhysRevB.109.205121,PhysRevLett.131.136502,PhysRevLett.131.136501,PhysRevB.107.L201109,PhysRevResearch.5.L032022,PhysRevB.108.245159,PhysRevB.109.045147,PhysRevLett.134.076503,PhysRevLett.134.066601,PhysRevLett.133.166503,PhysRevB.110.L161109,chen2025robust,PhysRevB.110.125142,kwan2024could,shen2025magnetorotonsmoirefractionalchern,PhysRevLett.131.136502,chen2025fractional,he2025fractional,li2025deep,luo2025solving,dong2024AHC1,zhou2024fractional,dong2024theorypentalayer,guo2024fractional,jonahMFCI2,kwan2023MFCI3,dong2024stability,soejima2024AHC2,tan2024parent,zeng2024sublattice,xie2024integerfractional,crepel2024efficientpredictionsuperlatticeanomalous,kudo2024quantumanomalousquantumspin,sarma2024thermal,huang2024selfconsistent,yu2024MFCI4,shavit2024entropy,huang2025displacement,huang2024impurityinducedthermalcrossoverfractional,guo2024beyondmeanfieldstudieswignercrystal,wei2025edge,tan2024wavefunctionapproachfractionalanomalous,zhou2024newclassesquantumanomalous,zeng2024berryphasedynamicssliding,li2025multiband,uzan2025hbnalignmentorientationcontrols,huo2025doesmoiremattercritical,PhysRevB.109.115116,PhysRevB.110.085120,PhysRevLett.133.066601,PhysRevLett.129.056804,PhysRevLett.133.206502,PhysRevLett.133.206504,PhysRevLett.133.206503,PhysRevB.112.075109,PhysRevX.14.041040,PhysRevB.110.205124,PhysRevB.110.205130}.

FQAH in $t\text{MoTe}_2$ feature a much richer interplay between topological order and symmetry breaking than their FQHE counterparts~\cite{PhysRevB.109.115116}. Unlike Landau levels where kinetic energy is quenched, FQAH emerges from moiré Chern bands that generally possess finite dispersion. This introduces a kinetic energy scale that competes with interactions, promoting a delicate intertwining between FQAH and proximate ordered states, such as charge density waves (CDWs)~\cite{cai2023signatures,zeng2023thermodynamic,park2023observation,xu2023observation,Redekop2024,ji2024local,xu2025signatures} or even unconventional superconductivity~\cite{xu2025signatures}. One particularly compelling theoretical proposal capturing this intertwining suggests that anyons can nucleate a halo of proximate CDW order~\cite{PhysRevB.110.085120}, rendering them extended objects with nontrivial internal structure and offering a new perspective on how FQAH give way to other competing phases. On the other hand, in moiré superlattice systems, transitions between FQAH and charge ordered phases are often driven by the displacement field, making it crucial for understanding the relationship between topological order and charge orders. One particularly intriguing scenario is the emergence of a QAHC~\cite{PhysRevB.109.115116,PhysRevLett.133.066601,chen2025fractional,PhysRevLett.129.056804,PhysRevLett.133.206502,PhysRevLett.133.206504,PhysRevLett.133.206503,PhysRevB.112.075109,PhysRevX.14.041040,PhysRevB.110.205124,PhysRevB.110.205130}, where the onset of charge order reshapes the topological response, converting a fractional Hall conductance into an integer one. Despite the fundamental importance of both anyon internal structures and displacement field effects, investigating these phenomena has been hindered by limited system size of exact diagonalization~\cite{li2021spontaneous,PhysRevB.111.125122,PhysRevB.108.085117,PhysRevLett.132.036501,xu2024maximally,PhysRevB.109.205121,PhysRevLett.131.136502,PhysRevLett.131.136501,PhysRevB.107.L201109,PhysRevResearch.5.L032022,PhysRevB.108.245159,PhysRevB.109.045147,PhysRevLett.134.076503,PhysRevLett.134.066601,PhysRevLett.133.166503,PhysRevB.110.L161109,chen2025robust,PhysRevB.110.125142,kwan2024could,shen2025magnetorotonsmoirefractionalchern}, and large-scale many-body simulations remain scarce~\cite{PhysRevLett.131.136502,chen2025fractional,he2025fractional,li2025deep,luo2025solving}.

In this paper, we directly address the intertwined FQAH-CDW physics by employing large-scale DMRG~\cite{white1992density,white1993density,SCHOLLWOCK201196} simulations on a minimal interacting lattice model, focusing on the paradigmatic moiré FQAH system of $3.7^\circ$ $t\text{MoTe}_2$ at filling $\nu=-2/3$. The real-space hopping parameters are obtained through Wannierization~\cite{devakul2021magic, PhysRevX.13.041026,xu2024maximally,crepel2024bridging} to accurately reproduce the continuum model band structure, while nearest-neighbor (NN) and next-nearest-neighbor (NNN) interactions, together with the displacement field, are treated as tunable parameters. We first demonstrate that robust FQAH ground state can already be stabilized in a minimal model with NN interactions only. By introducing single particle doping into the FQAH ground state, we obtain clear numerical evidence for fractionalized anyon excitations with density-wave halo~\cite{PhysRevB.110.085120} internal structure. By further including NNN interactions and the displacement field, we map out the resulting phase diagram, revealing that both QAHC and topologically trivial CDW phases can be stabilized. Our findings establish a concrete microscopic framework for understanding the intertwined topological and symmetry breaking physics in moiré superlattice systems.

\begin{figure}[t]
    \centering
    \includegraphics[width=0.8\linewidth]{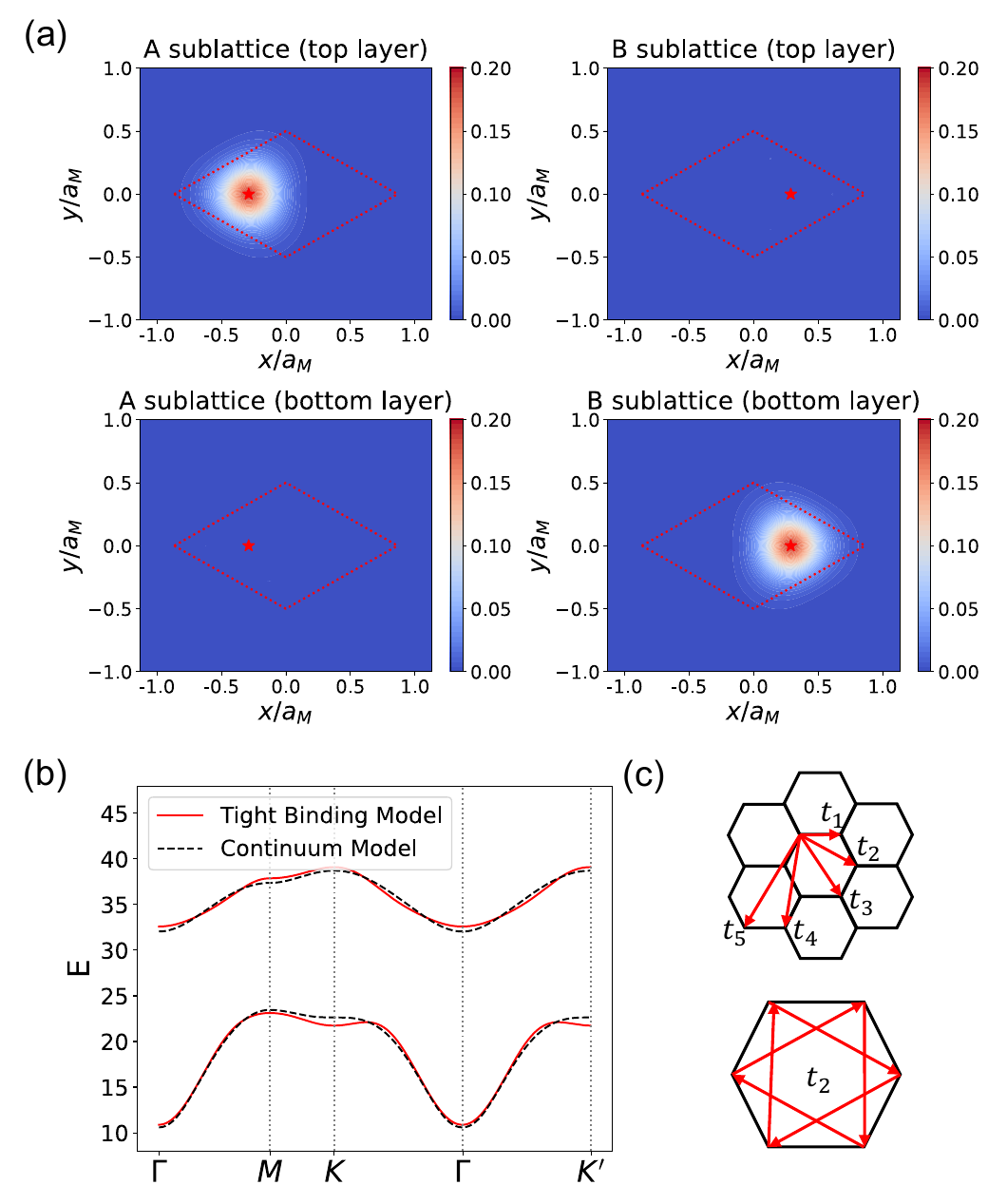}
    \caption{(a) Density distribution of the Wannier functions, with Wannier centers indicated by stars and moiré unit cell shown as dashed line. (b) Comparison of the band structure between the tight-binding model and the continuum model. (c) Illustration of the hopping terms in tight-binding model on a honeycomb lattice. The next nearest neighbor hopping $t_2$ carry phase $+2\pi/3$ along the indicated arrow direction.}
    \label{Fig1}
\end{figure}

\textit{Model and methods.}---
Our starting point is the continuum model~\cite{PhysRevLett.122.086402} for $3.7^\circ$ $t\text{MoTe}_2$, which is typically formulated in momentum space and has been widely used to study the moiré band structure and many-body properties. However, this formulation is not directly compatible with many-body methods like DMRG, which generally require a real space tight-binding representation.
To bridge this gap, we construct Wannier orbitals via Wannierization~\cite{supp}, and we restrict to spinless fermion model due to the robust experimental evidence of spin-valley polarized ferromagnetism~\cite{anderson2023programming}. Since the topmost moiré valence band carries a non-zero Chern number ($C=-1$), we incorporate the second valence band ($C=+1$) to eliminate the topological Wannier obstruction.
Leveraging the layer-polarization inherent to stacking of transition metal dichalcogenides, we can obtain two exponentially localized Wannier functions in a more convenient way~\cite{devakul2021magic, PhysRevX.13.041026,xu2024maximally,crepel2024bridging,supp}. 
The resulting two orbitals are shown in Fig.~\ref{Fig1}(a), where they are localized around the Wannier centers $(\pm a_M/2\sqrt{3},0)$, forming a honeycomb lattice, with A orbital predominantly on the top layer and B orbital on the bottom layer, as expected.

With the Wannier functions, we construct a real space tight-binding model:
\begin{equation}
    H_0 = \sum_{ij}t_{ij} c_i^\dagger c_j 
\end{equation}
where $t_{ij}$ denotes the hopping amplitude between sites $i$ and $j$, and $c_i^\dagger$ is the creation operator for a spinless fermion at $K$ valley. By retaining up to the fifth-nearest-neighbor hopping with parameters $(t_1, t_2, t_3, t_4, t_5) = (-3.88, 1.93 e^{i2\pi/3}, 0.94, -0.34, -0.13)$ meV (see hopping patterns in Fig.~\ref{Fig1}(c)), we obtain a tight-binding model that takes the form of a generalized Haldane model~\cite{PhysRevLett.61.2015} and faithfully reproduces the dispersion of the continuum model, as shown in Fig.~\ref{Fig1}(b).

Next, we address the interaction effects and the influence of the displacement field. In contrast to the single-particle band structure, determining the precise form and strength of interactions is challenging due to the complex screening environment in moiré systems. Therefore, we focus on a minimal model where only the NN and NNN interactions are retained:
\begin{equation}
    H_\text{int} = V_1\sum_{\langle ij\rangle} n_i n_j + V_2\sum_{\langle\langle ij\rangle\rangle} n_i n_j
\end{equation}
where $V_1, V_2$ are the strength of NN and NNN interaction, and $n_i$ represent the density at site $i$. This approximation is justified when the screening effects in $t\text{MoTe}_2$ are strong enough to suppress longer-range components. Moreover, due to the layer polarization of the Wannier functions, the displacement field effects can be captured as a sublattice staggered potential:
\begin{equation}
    H_D = D \sum_{i} (-1)^i n_i,
\end{equation}
where $(-1)^i=\pm 1$ for $i\in A/B$ sublattice, $D$ denotes the strength of the displacement field. The full interacting Hamiltonian is thus given as $H = H_0 + H_\text{int}+H_D$.

To investigate the interacting physics of the model, we employ the DMRG algorithm, which is well suited for studying the interplay between topology and strong correlations. In our simulations, we exploit $U(1)$ charge conservation symmetry, with the total particle number is chosen to ensure the correct bulk filling $\nu = -2/3$.
We adopt $L_y=6$ cylinder geometry, and keep bond dimensions up to $\chi=4000$, which is sufficient to accurately extract the ground state properties of the system.

\begin{figure}[t]
    \centering
    \includegraphics[width=1\linewidth]{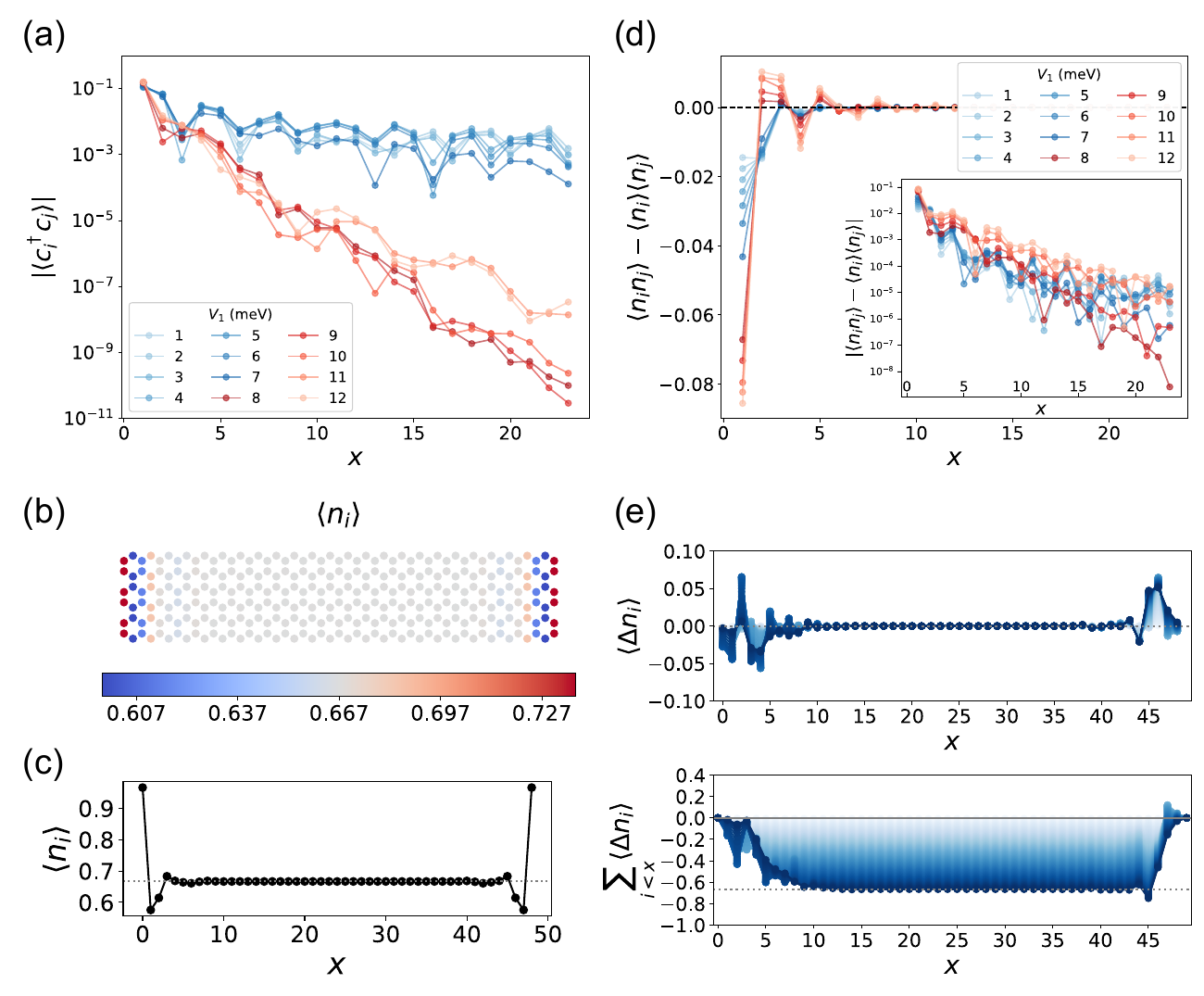}
    \caption{(a) Single particle Green's functions for various NN interaction $V_1$. (b) Real space density profile $\langle n_i\rangle$ at $V_1=8$ meV. (c) The same density distribution averaged along $y$ direction. (d) Connected density correlations for different NN interaction $V_1$ in linear (semi-log) scale for main (inset) figure. (e) Adiabatic charge pumping under insertion of $2\pi$ flux thread through the cylinder at $V_1=8$ meV. The upper panel shows the local density change, while the lower panel shows its cumulative sum from the left boundary, where the color shading indicates the magnitudes of the inserted flux.}
    \label{Fig2}
\end{figure}

\textit{FQAH state.}---We begin by examining whether a minimal model of $t\text{MoTe}_2$ can realize a FQAH phase at $\nu=-2/3$, focusing on small to intermediate NN repulsion $V_1$ regime, while keeping both NNN repulsion $V_2$ and the displacement field $D$ at zero. Fig.~\ref{Fig2}(a) illustrates the single particle Green's functions $\langle c^\dagger_i c_j\rangle$, which reveal a sharp qualitative distinction between the weak and intermediate interaction regimes. For $V_1<8$ meV, the Green's functions decay algebraically, a characteristic of Fermi liquid behavior. In contrast, for $V_1\geq 8$ meV, it exhibits a rapid exponential decay, indicating the onset of an insulating state with a finite single particle gap. 

To elucidate the nature of this insulator, we examine the real-space density profile for a representative case at $V_1=8$ meV, as shown in Fig.~\ref{Fig2}(b) and (c). Except for the boundary, the bulk density is pinned exactly at $\langle n \rangle=2/3$, consistent with the featureless, translationally invariant nature expected for a FQAH state. Fig.~\ref{Fig2}(d) further demonstrates the connected density correlations $\langle n_in_j\rangle - \langle n_i \rangle\langle n_j\rangle$ for different interaction $V_1$. The inset reveals that the decay behavior changes qualitatively from algebraic ($V_1<8$ meV) to exponential ($V_1\geq 8$ meV), consistent with a transition into fully gapped state. Moreover, the main panel highlights that the NN density correlations are strongly suppressed in the insulating regime, reflecting local electron avoidance and the formation of a correlation hole, consistent with FQAH physics.

While the density profile and correlations strongly suggest a FQAH phase, a definitive confirmation requires direct probing the topological properties. The DMRG simulations on cylinder geometry offer a natural framework for accessing Hall conductance of the system through Laughlin's adiabatic charge pumping via flux insertion~\cite{PhysRevB.23.5632,thouless1982quantized,PhysRevB.31.3372,PhysRevLett.110.236801,supp}. The presence of nearby competing states makes a sufficiently slow flux insertion essential. Accordingly, we employ a system with length $L_x=49$ and discretize the $2\pi$ flux insertion into $100$ steps, while retaining the bond dimension to $\chi=1000$ to maintain computational feasibility. As shown in Fig.~\ref{Fig2}(e), for the $V_1=8$ meV system, adiabatically threading a $2\pi$ flux through the cylinder induces density changes localized at the edges, while leaving the bulk density invariant. This process results in a net charge transfer of exactly $\Delta Q = 2/3$ from the left edge to the right edge, corresponding to a fractionally quantized Hall conductance of $\sigma_{xy}=-2/3$. Taken together, these numerical observations provide unambiguous evidence that our $t\text{MoTe}_2$ model, even with only NN repulsion $V_1$, hosts a robust FQAH ground state.

\begin{figure}[t]
    \centering
    \includegraphics[width=0.8\linewidth]{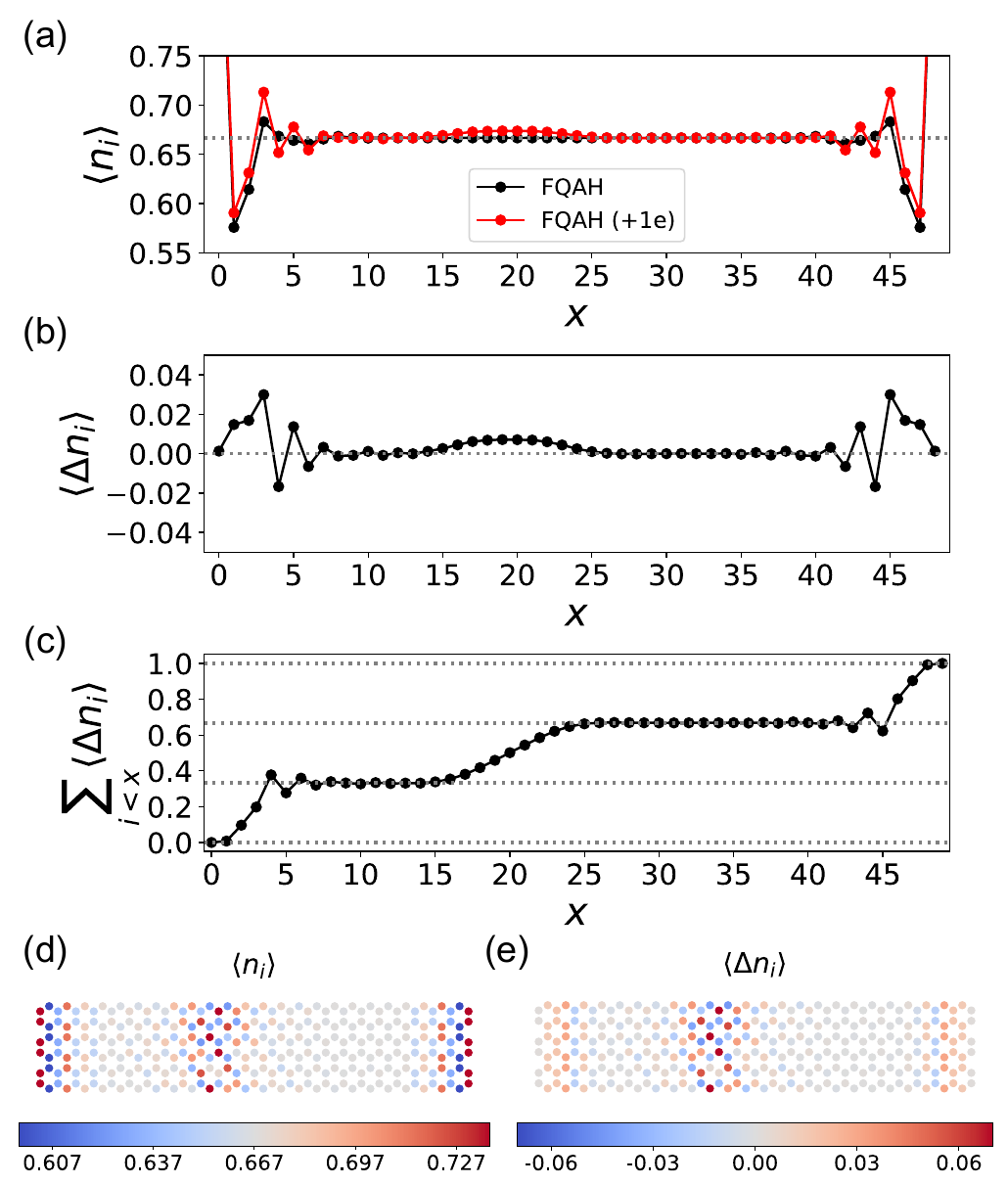}
    \caption{(a) $y$-direction averaged real space density profile for the FQAH at $V_1=8$ meV (black) and for the system with one added electron (red). (b) Density difference $\langle \Delta n_i\rangle$ between these two systems. (c) Cumulative sum of the density difference from the left boundary. The gray dashed lines at $0, 1/3, 2/3, 1$ mark the expected fractionally quantized value of charge. (d) Real space density distribution for the added-electron system in (a). (e) Real space distribution for the density difference in (b). }
    \label{Fig3}
\end{figure}

\textit{Anyon density-wave halo.}---Beyond the fractionally quantized Hall conductance, another definitive hallmark of intrinsic topological order of FQAH phase is the existence of fractionalized anyon excitations. Our DMRG simulations directly demonstrate this fractionalization by introducing a single electron into the FQAH ground state. Fig.~\ref{Fig3}(a) presents the density distribution for the FQAH state at $V_1=8$ meV and for the system with one extra electron, and Fig.~\ref{Fig3}(b) shows the resulting density difference. These results clearly reveal that the added single electron charge separates into three spatially distinct regions, two located at the boundaries and one inside the bulk, which is indicative of three anyon excitations.
To unambiguously verify the fractional nature of these excitations, we calculate the cumulative sum of the density difference starting from the left boundary, as shown in Fig.~\ref{Fig3}(c). This profile clearly demonstrates that each region carries precisely $1/3$ of the electron charge, consistent with the expected fractionalization of FQAH at $\nu=-2/3$. These findings confirm that doping the FQAH system injects fractionalized anyons rather than electrons into the system.

The ability to spatially isolate a single anyon provides a unique opportunity to numerically resolve its internal structure. Fig.~\ref{Fig3}(d) and (e) show the real space density for the doped system and the corresponding density difference associated with the anyon charge distribution. Crucially, the anyons exhibit strong real-space density modulations, consistent with the recently proposed anyon density-wave halo picture~\cite{PhysRevB.110.085120}. In this framework, the internal modulation reflects the proximity to a charge ordered phase. Specifically, if the FQAH state is viewed as a condensate of composite bosons, the anyon corresponds to a vortex defect. Within the vortex core, the condensate amplitude is suppressed, allowing the competing density-wave order to manifest. We further demonstrate that such anyon density-wave halo persists for finite NNN interaction $V_2$ and displacement field $D$ with a much cleaner charge order pattern~\cite{supp}. This phenomenon highlights the intricate interplay between the FQAH phase and the proximate charge ordered phases.

\begin{figure}[t]
    \centering
    \includegraphics[width=0.9\linewidth]{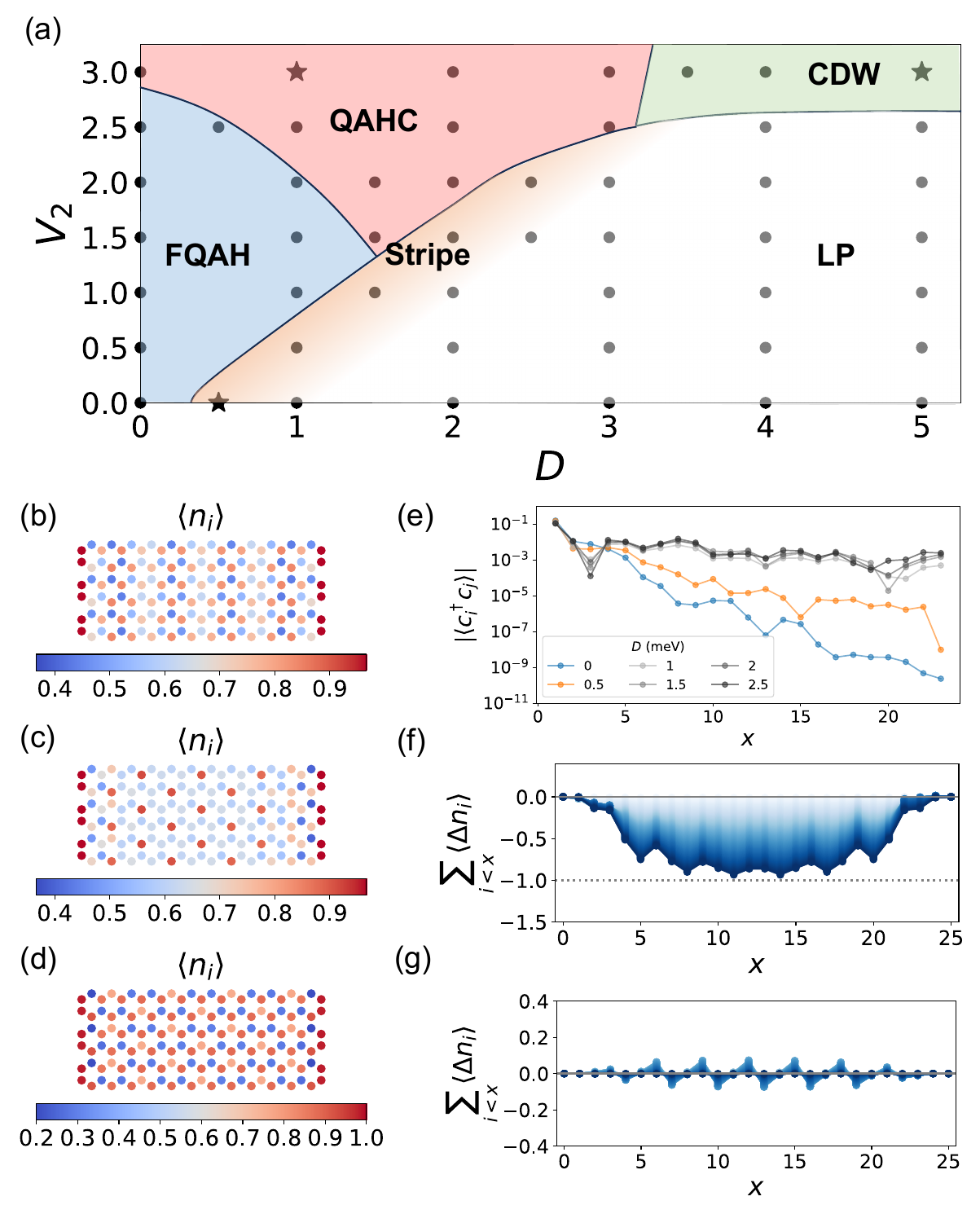}
    \caption{(a) Ground state phase diagram as a function of NNN interaction $V_2$ and displacement  field $D$ at fixed NN interaction $V_1=10$ meV, obtained using $L_x=25$ cylinders. The stripe region is depicted only schematically by an orange shade. (b-d) Real space density distribution $\langle n_i\rangle$ for three representative systems in different charge ordered phases marked by stars in the phase diagram: (b) $V_2=0$ meV, $D=0.5$ meV; (c) $V_2=3$ meV, $D=1$ meV; (d) $V_2=3$ meV, $D=5$ meV. (e) Single particle Green's functions $\langle c_i^\dagger c_j\rangle$ for various displacement fields $D$ at $V_1=10$ meV and $V_2=0$ meV, obtained from $L_x=49$ cylinders. (f, g) Adiabatic charge pumping under insertion of $2\pi$ flux for the same representative systems as in (c) and (d), where the color shading indicates the magnitudes of the inserted flux. }
    \label{Fig4}
\end{figure}

\textit{Displacement field effects.}---We now turn to the role of the displacement field in $t\text{MoTe}_2$, which drives transitions between the FQAH and competing charge ordered phases. By keeping NN repulsion at a representative value of $V_1=10$ meV and additionally incorporating the NNN repulsion $V_2$, we uncover a rich landscape of ground states, as summarized in Fig.~\ref{Fig4}(a), hosting the FQAH phase, two $\sqrt{3}\times\sqrt{3}$ CDW phases with distinct topological character (i.e. QAHC and CDW), a stripe phase, and a layer polarized (LP) phase. 
Within this phase diagram, the FQAH phase identified previously remains stable against finite NNN interaction $V_2$ and displacement field $D$, as expected from the robustness of the fully gapped topological order. While a finite displacement field induces weak sublattice staggered density modulations, it does not immediately destroy the topological order. Notably, the inclusion of weak NNN repulsion $V_2$ serves to further stabilize the FQAH phase.

In the regime of weak $V_2$, increasing the displacement field $D$ drives a transition from the FQAH phase into a stripe phase. A representative real space density distribution for the stripe phase is shown in Fig.~\ref{Fig4}(b), which exhibits unidirectional charge modulation with a period of approximately three unit cells along the $x$-direction. With further increase of $D$, the stripe pattern melts, and the system eventually transitions into a LP phase, where the electrons are almost entirely confined to a single sublattice in response to the strong displacement field. To shed light on the transport properties of these phases, Fig.~\ref{Fig4}(e) shows the single particle Green’s function at $V_2 = 0$ meV for various displacement fields, revealing insulating behavior in the stripe phase and metallic behavior in the LP phase. 
This metallic behavior can be understood from an effective model on the hole-rich sublattice in the large displacement field limit, where the NNN interaction $V_2$ is insufficient to destabilize the Fermi liquid behavior.

For sufficiently large $V_2$, the system favors the formation of $\sqrt{3}\times\sqrt{3}$ CDW phases. In our simulations, we identify two qualitatively distinct CDW patterns, with representative real-space density distributions presented in Fig.~\ref{Fig4}(c) and (d), and the transition between them appears to be first order. Although these two CDW phases are not distinguished by symmetry, they differ in whether the spontaneous translational symmetry breaking predominantly occurs on the hole-poor or hole-rich sublattice.
Moreover, adiabatic charge pumping simulations indicate that the two CDW phases differ qualitatively in their topological properties. Fig.~\ref{Fig4}(f) shows the charge pumping at the representative system with $V_2=3$ meV and $D=1$ meV, where adiabatically threading a $2\pi$ flux through the cylinder induces an integer net charge transfer $\Delta Q = 1$ from the left edge to the right edge, demonstrating that the system realizes a QAHC phase in this parameter regime with both CDW order and integer quantized Hall conductance $\sigma_{xy}=-1$.
However, performing the same adiabatic charge pumping procedure at $V_2=3$ meV and $D=5$ meV shows no evidence of net charge transfer, suggesting a topologically trivial CDW phase.
Taken together, the displacement field serves as an indispensable tuning parameter that gives rise to a rich set of ground states in $t\text{MoTe}_2$.

\textit{Summary and outlook.}---In this paper, we present a comprehensive microscopic study of the interplay between FQAH and charge orders in $t\text{MoTe}_2$. By performing large-scale DMRG simulations on a minimal interacting lattice model, we provide clear numerical evidences for robust FQAH ground state, anyon density-wave halo, and a rich displacement-field tunable ground state landscape. These results provide valuable insight from both theoretical and experimental perspectives.

One central implication of our results is that, FQAH and charge order in $t\text{MoTe}_2$ do not merely compete energetically, but intertwine non-trivially via the internal structure of excitations. This observation aligns remarkably well with the anyon density-wave halo picture proposed recently by Song and Senthil~\cite{PhysRevB.110.085120}. Such intertwining nontrivially reshapes the low-energy effective theory and has far-reaching implications. One prominent consequence is that, defects in the nearby CDW phase can carry fractional charge and orbital magnetization, reflecting the proximity to the underlying FQAH phase, which deserves future numerical and experimental verification.

Experimentally, our ground state phase diagram on displacement field effects (Fig.~\ref{Fig4}(a)) has direct relevance for interpreting the recent observations in $t\text{MoTe}_2$. 
Near the $\nu=-2/3$ FQAH state, recent experiments have reported a pronounced anomalous Hall response and several reentrant integer quantum anomalous Hall states as the filling factor and displacement field are tuned~\cite{xu2025signatures}. This phenomenology is naturally related to the broad QAHC region in our phase diagram. An important open question is whether such QAHC with integer quantized Hall conductance can be stabilized exactly at the fractional filling $\nu=-2/3$ in future experiments of $t\text{MoTe}_2$. At larger displacement fields, our phase diagram suggests a topologically trivial CDW state, consistent with the insulating behavior observed experimentally.
Looking forward, another intriguing open question is whether the unconventional superconductivity recently observed in $t\text{MoTe}_2$~\cite{xu2025signatures} can be understood within this or related microscopic frameworks~\cite{PhysRevLett.60.2677,PhysRevLett.63.903,doi:10.1073/pnas.2426680122,wang2025chiralsuperconductivitynearfractional,kuhlenkamp2025robustsuperconductivitydopingchiral,Pichler_2025,nosov2025anyonsuperconductivityplateautransitions,Shi_2025,shi2025anyondelocalizationtransitionsdisordered,shi2025nonabeliantopologicalsuperconductivitymelting,guerci2025fractionalizationchiraltopologicalsuperconductivity}. 
We believe our work will stimulate further theoretical and experimental interest in investigating the intertwined nature of correlated topological phases in moiré systems.

\textit{Acknowledgments.}---We thank Xue-Yang Song for helpful discussions. This work is supported in part by MOSTC under Grant No. 2021YFA1400100 (H.Y.), by NSFC under Grant Nos. 12347107 (C.T., M.-R.L., H.Y.) and 12334003 (H.Y.), and by the New Cornerstone Science Foundation through the Xplorer Prize (H.Y.).

\bibliography{ref.bib}

\end{document}


\title{Supplemental material for ‘Fractional quantum anomalous Hall and anyon density-wave halo in a minimal interacting lattice model of twisted bilayer MoTe$_2$’}

\author{Chuyi Tuo}
\affiliation{Institute for Advanced Study, Tsinghua University, Beijing 100084, China}
\author{Ming-Rui Li}
\affiliation{Institute for Advanced Study, Tsinghua University, Beijing 100084, China}
\author{Hong Yao}
\email{yaohong@tsinghua.edu.cn}
\affiliation{Institute for Advanced Study, Tsinghua University, Beijing 100084, China}

\date{\today}
\maketitle

\section{Construction of the Wannier functions}
In this section, we provide details of the procedure for constructing the Wannier functions. 
Our starting point is the continuum model~\cite{PhysRevLett.122.086402} for $t\text{MoTe}_2$, where the Hamiltonian for $K$ valley is given by:
\begin{equation}
H_K(\bm{r})=\left(\begin{array}{cc}
-\frac{\hbar^2\left(\bm{k}-\boldsymbol{\kappa}_{+}\right)^2}{2 m^*}+\Delta_+(\bm{r}) & \Delta_{\text{T}}(\bm{r}) \\
\Delta_{\text{T}}^{\dagger}(\bm{r}) & -\frac{\hbar^2\left(\bm{k}-\boldsymbol{\kappa}_{-}\right)^2}{2 m^*}+\Delta_-(\bm{r})
\end{array}\right).
\end{equation}
Here $m^*=0.6m_e$ is the effective mass of monolayer MoTe$_2$, $\bm{\kappa}_\pm$ denote the wave vectors of the mini Brillouin zone corners. The moiré potential $\Delta_\pm(\bm{r})$ and interlayer tunneling $\Delta_{\text{T}}(\bm{r})$ are given by:
\begin{align}
\Delta_{\pm}(\bm{r}) &= 2 v \sum_{j=1,3,5} \cos \left(\bm{g}_j \cdot \bm{r} \pm \psi\right)\\
\Delta_{\text{T}}(\bm{r}) &= w\left(1+e^{-i \bm{g}_2 \cdot \bm{r}}+e^{-i \bm{g}_3 \cdot \bm{r}}\right)
\end{align}
and the moiré wave vectors $\bm{g}_j$ are obtained by rotation of $\bm{g}_1 = (\frac{4\pi}{\sqrt{3}a_M}, 0)$ by $(j-1)\pi/3$. The continuum model parameters are obtained by large scale ab initio simulations $(v, \psi, w) = (20.8 \text{meV}, 107.7^\circ, -23.8 \text{meV})$~\cite{PhysRevLett.132.036501}. Here we focus on the experimentally relevant twist angle $\theta=3.7^\circ$, and the moiré lattice constant is given by $a_M \approx a_0/\theta$ with $a_0 = 3.52$\AA. Motivated by the robust experimental evidence for spin-valley polarization ferromagnetism~\cite{anderson2023programming} in the FQAH regime of interest, we restrict our attention to consider spinless fermion at $K$ valley in this paper. And, as stated in the main text, we consider the top two moiré valence bands to eliminate the topological Wannier obstruction.

We construct exponentially localized Wannier functions in a particularly convenient manner~\cite{devakul2021magic, PhysRevX.13.041026,xu2024maximally,crepel2024bridging} by exploiting the layer polarization properties of $t\text{MoTe}_2$. In our case of $3.7^\circ$ $t\text{MoTe}_2$, the Wannier centers are located at XM and MX regions, and the associated Wannier functions are polarized on opposite layers as a consequence of the large metal-chalcogenide chemical potential difference. The Wannierization procedure utilizing the layer polarization properties is described as follows: (i) diagonalize the continuum model in momentum space to obtain the Bloch eigenstates $\ket{\phi_{n\bm{k}}}$, where $n=1,2$ is the band index; (ii) construct the layer polarization operator $P_{nn'}(\bm{k}) = \bra{\phi_{n\bm{k}}} \tau^z \ket{\phi_{n'\bm{k}}}$, where $\tau^z $ denotes the $z$ component of the layer pesudospin. (iii) diagonalize the layer polarization operator $P_{nn'}(\bm{k})$ at each momentum $\bm{k}$ via a unitary transformation $U_{nn'}(\bm{k})$, and the resulting two eigenstates $\ket{\tilde{\phi}_{n\bm{k}}}=\sum_{n'} U_{nn'}(\bm{k}) \ket{\phi_{n'\bm{k}}}$ are predominantly polarized on opposite layers. (iv) fix the gauge of each $\ket{\tilde{\phi}_{n\bm{k}}}$ by imposing it to be real and positive at the corresponding Wannier centers (i.e. XM or MX regions $(\pm a_M/2\sqrt{3},0)$). (v) the Wannier functions at lattice vector $\bm{R}$ can then be constructed by $\ket{W_{n\bm{R}}} = \tfrac{1}{\sqrt{N}} \sum_{\bm{k}} e^{-i\bm{k}\cdot \bm{R}} \ket{\tilde{\phi}_{n\bm{k}}} $, where $N$ is the total number of unit cells. The resulting Wannier functions are illustrated in Fig.~1(a) in the main text, which are well exponentially localized and display the correct layer polarization and Wannier center locations.

\section{Adiabatic charge pumping}
In this section, we provide additional details on the adiabatic charge pumping method~\cite{PhysRevB.23.5632,thouless1982quantized,PhysRevB.31.3372,PhysRevLett.110.236801} adopted in the main text, which serves as an indispensable numerical tool for accessing the topological properties of the many-body ground state in our DMRG simulations on cylinder geometry. In simple terms, adiabatically threading flux $\Phi$ through the cylinder induces an electric field along the circumferential direction, which the Hall response converts into a transverse current along the cylinder axis, resulting in a net charge transfer between the two open edges in proportional to the Hall conductance $\sigma_{xy}$. The quantization of the Hall conductance $\sigma_{xy}$ in a gapped phase can also be understood straightforwardly from gauge invariance under insertion of integer number of $2\pi$ flux quanta.

Practically, flux insertion is implemented by imposing twisted boundary conditions around the circumference of the cylinder, i.e., $c_{x,y+L_y}^\dagger = e^{i\Phi} c_{x,y}^\dagger$ when writing in terms of the brick-wall coordinate of the honeycomb lattice. Therefore, the hopping terms acquire a phase $+\Phi$ when crossing the circumferential boundary in the positive direction, and a phase $-\Phi$ when crossing in the opposite direction, while other terms in the Hamiltonian remain invariant. 
In our DMRG simulations, adiabatic charge pumping is implemented as follows:
(i) the $2\pi$ flux insertion is discretized into $M$ steps, where $\Phi_m = 2\pi m/M$ with $m=0,\dots,M$;
(ii) for each flux value $\Phi_m$, we compute the many-body ground state of the Hamiltonian $H(\Phi_m)$, using the converged ground state at $\Phi_{m-1}$ as the initial state to ensure adiabatic continuity;
(iii) at each $\Phi_m$, we measure the real-space density profile to extract the pumped charge.
In our simulations, since the ground state is intertwined with charge orders, a sufficiently slow flux insertion (i.e., a relatively large $M$) is required to ensure numerical reliability.

\begin{figure}[H]
    \centering
    \includegraphics[width=1\linewidth]{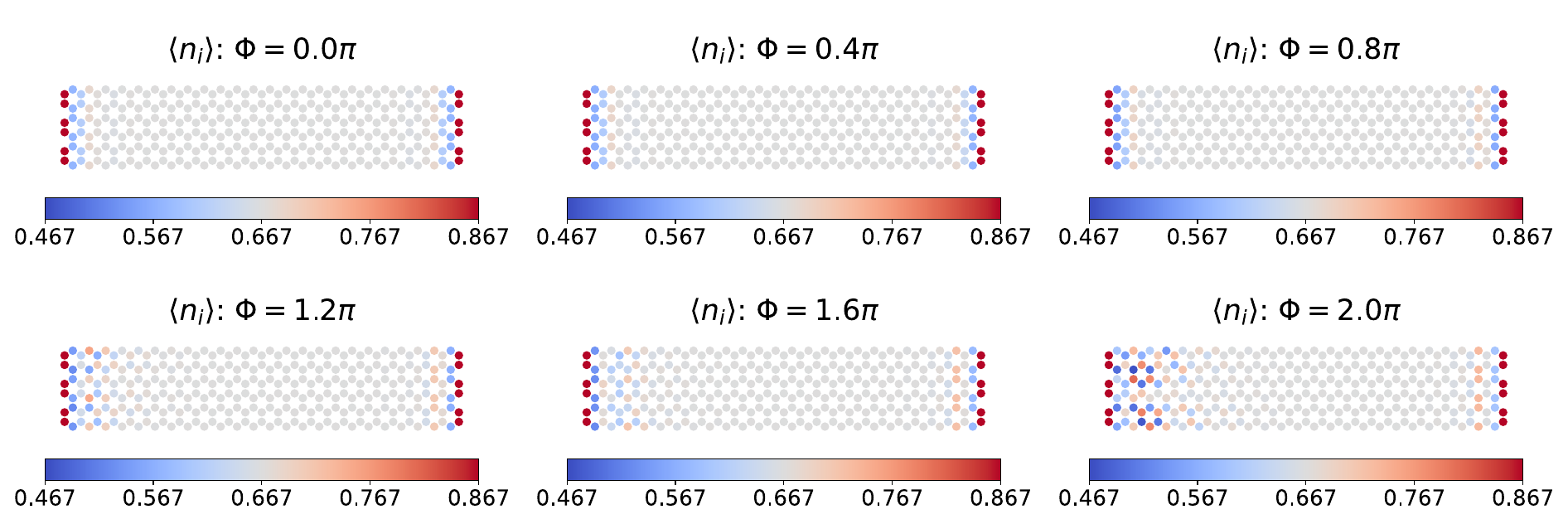}
    \caption{Density distributions for representative flux values in the FQAH phase with $V_1=8$ meV.}
    \label{FQAH_pump}
\end{figure}

\begin{figure}[H]
    \centering
    \includegraphics[width=0.8\linewidth]{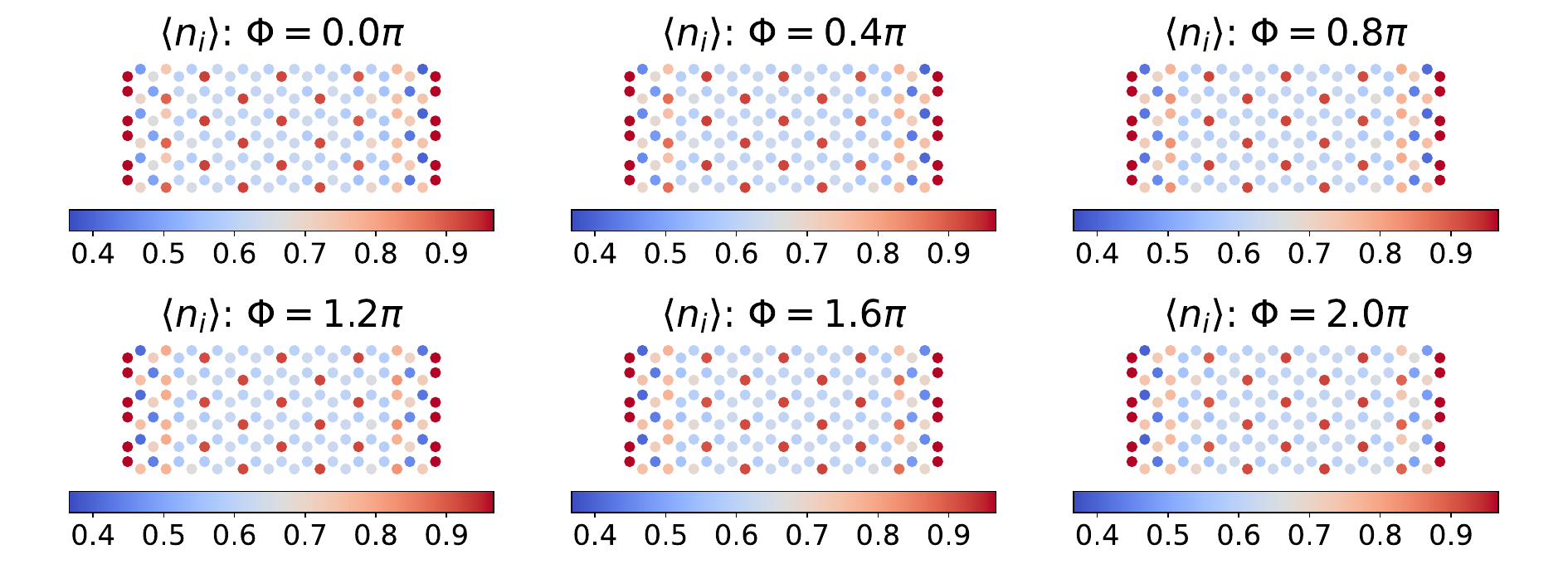}
    \caption{Density distributions for representative flux values in the QAHC phase with $V_1=10$ meV, $V_2=3$ meV, $D=1$ meV.}
    \label{QAHC_pump}
\end{figure}

\begin{figure}[H]
    \centering
    \includegraphics[width=0.8\linewidth]{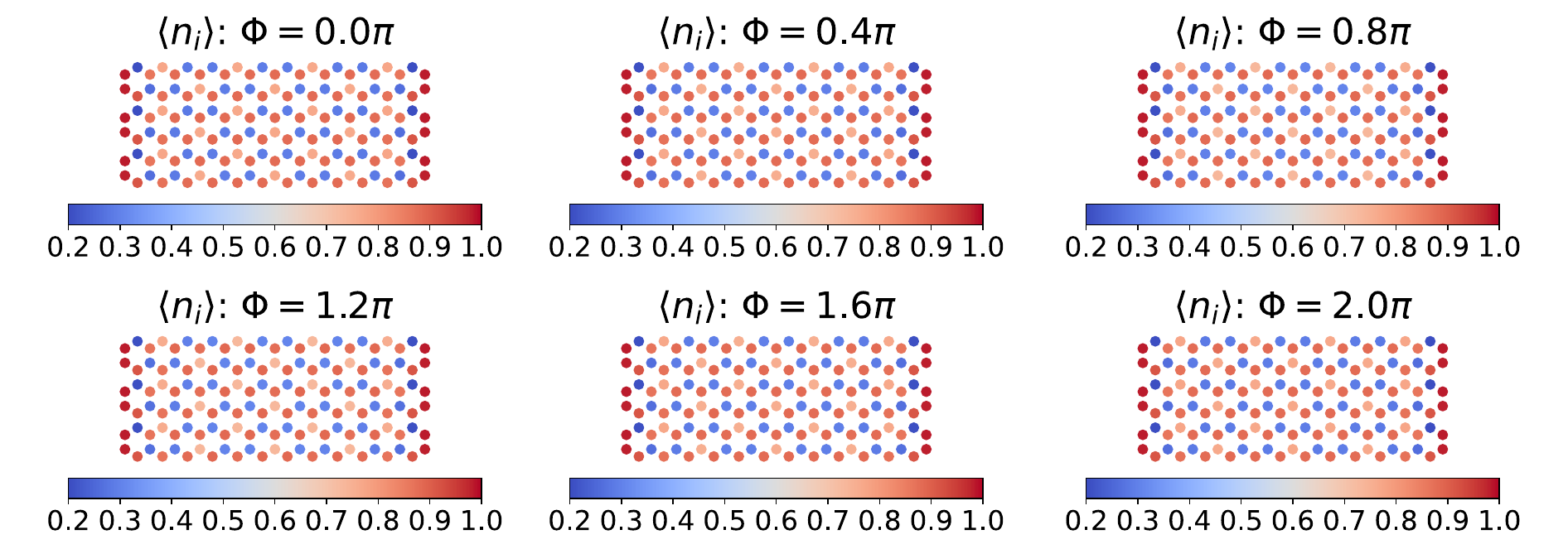}
    \caption{Density distributions for representative flux values in the CDW phase with $V_1=10$ meV, $V_2=3$ meV, $D=5$ meV.}
    \label{CDW_pump}
\end{figure}

In the main text, we have provided robust numerical evidences for the quantization of Hall conductance $\sigma_{xy}$ in various phases. Here, we present additional real space information associated with these adiabatic charge pumping processes. Figures~\ref{FQAH_pump} and \ref{QAHC_pump} illustrate the adiabatic charge pumping for the FQAH and QAHC phases with several representative values of the flux $\Phi$, showing that the density changes occur predominantly at the edges while the bulk remains stable. In contrast, Fig.~\ref{CDW_pump} illustrates that the topologically trivial CDW phase exhibits almost no response to the inserted flux. These results suggest that adiabatic charge pumping serves as a powerful and direct numerical probe of the topological properties for the many-body ground state.

\section{Anyon density-wave halo close to the FQAH-QAHC transition}
\begin{figure}[H]
    \centering
    \includegraphics[width=0.8\linewidth]{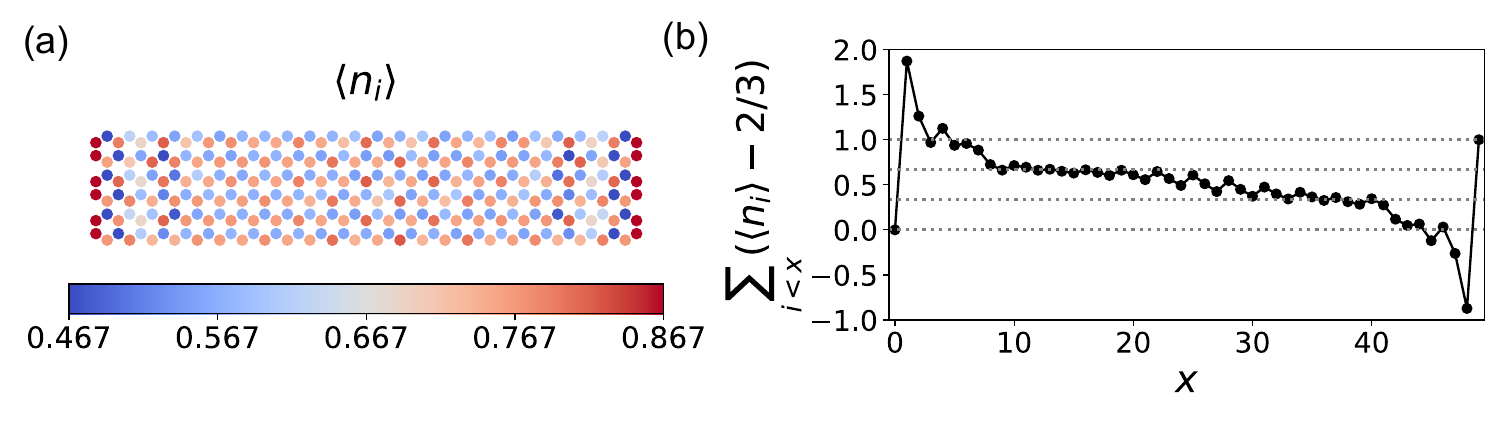}
    \caption{(a) Density distribution for system with $V_1=10$ meV, $V_2=1.5$ meV, $D=1$ meV, showing a clear anyon density-wave halo with $\sqrt{3}\times\sqrt{3}$ CDW pattern. (b) Cumulative sum of the density deviation $\langle n_i\rangle-2/3$ from the left boundary, and the gray dashed lines mark values of $0,1/3,2/3,1$.}
    \label{halo}
\end{figure}

In the main text, we have demonstrated that anyon density-wave halo~\cite{PhysRevB.110.085120} can occur in systems with only NN interaction $V_1$. Here, we provide additional numerical evidence that such phenomena also persist in the presence of finite NNN interaction $V_2$ and displacement field $D$ close to the FQAH-QAHC transition. Fig.~\ref{halo}(a) illustrates the real space density distribution $\langle n_i \rangle$ for a representative system close to the FQAH-QAHC transition at $V_1=10$ meV, $V_2=1.5$ meV, $D=1$ meV, where an additional electron relative to the $\nu=-2/3$ filling is included to compensate for the open boundary effects. While most of the bulk shows sublattice staggered density modulation induced by the displacement field, the central region of the bulk develops $\sqrt{3}\times\sqrt{3}$ CDW pattern similar to the QAHC phase (see Fig.~4(c) in the main text), which is consistent with an anyon density-wave halo. To obtain decisive evidence that this density modulation is associated with an anyon, we compute the cumulative sum of the density deviation $\langle n_i\rangle-2/3$ measured from the left boundary, as shown in Fig.~\ref{halo}(b). From this cumulative sum, we find that the extra electron fractionalizes, resulting in a net charge of $-1/3$ associated with the anyon density-wave halo in the bulk, as well as an excess charge of $2/3$ at each open edge. These results provide strong evidence that anyon density-wave halos persist over a broad parameter regime in $t\text{MoTe}_2$ and encode the characteristics of nearby charge ordered phases, highlighting the close interplay between the FQAH state and proximate ordered phases.

\bibliography{ref.bib}